# interface
## *electronic chamber ensemble*

**Curtis Bahn**
Integrated Electronic Arts, iEAR Studios
DCC 135, Renssealaer Polytechnic Institute
110 8th Street, Troy, NY 12180-3590
(518) 276-4032
crb@rpi.edu

**Dan Trueman**
Computer Music Center
Columbia University
632 w.125th Street, New York, NY 10027
(212) 854-9266
dan@music.columbia.edu

**ABSTRACT**
This paper presents the interface developments and music of the duo "interface," formed by Curtis Bahn and Dan Trueman. We describe gestural instrument design, interactive performance interfaces for improvisational music, spherical speakers (multi-channel, outward-radiating geodesic speaker arrays) and Sensor-Speaker-Arrays (SenSAs: combinations of various sensor devices with spherical speaker arrays). We discuss the concept, design and construction of these systems, and, give examples from several new published CDs of work by Bahn and Trueman.

**Keywords**
Interactive Music Performance, Gestural Interface, Sonic Display, Sensor/Speaker Array, SenSA.

**INTRODUCTION**

"interface" is an improvisatory electronic music duo formed in 1995 by composers Curtis Bahn and Dan Trueman focusing on issues of spontaneous musical interaction, physical gesture, and sonic display in the performance of live-electronic "chamber music." We have created extended sensor-based instrumental interfaces, new movement interfaces for dance, complex interactive computer music performance environments, and a family of multi-channel spherical speaker arrays. These enhanced musical devices have fundamentally changed the way in which we interact in performance, both physically and sonically, extending traditional chamber music practice in a technological context.

We consider our entire systems, from physical instruments, sensor interfaces, interactive computer music environments to spherical speaker arrays, to be both extended instruments and non-linear compositions: composed instruments. The combination of physical interface and sonic reinforcement provides aural and tactile feedback cues that are essential to our music making.

These interfaces extend and abstract traditional approaches to live musical performance, and allow for a direct physicality and musical gesture to be communicated in electronic music. Our approach privileges the group's interaction, or an individual's musical expression, over "interaction" with the computer, re-enforcing the activity and social context of music making.

This paper covers a body of work from the period of 1998-2000 in which we developed numerous new instruments, interfaces and sonic display devices culminating in several significant public performances and a live CD, "./swank," released on the new Cycling74 label, C74.

**COMPOSED INSTRUMENTS**

Both Bahn and Trueman grew up playing string instruments in various contexts from orchestras and acoustic chamber ensembles to electric rock and jazz bands. "Interface" began as a purely acoustic duo, gradually integrating electronics into the ensemble over a period of several years. Their primary performance interfaces are still built around their original string instruments, now used sometimes simply as subtle computer controllers. Covered here are their main performance instruments: the Sensor Bass—Sbass—and the Rbow.

### Rbow - Sensor Bow

The Rbow, constructed by Trueman and Perry Cook, consists of a traditional violin bow with motion sensors (a biaxial accelerometer, mounted at the frog) and pressure sensors (mounted between the hair and the stick in two locations). It can be played by itself, using the shoulder or other surface as a point of resistance, or on any violin. Trueman uses it primarily with a six-string, solid-body electric violin, and in combination with pitch, amplitude, and overtone detection of the electric violin signal.

The Rbow was motivated by Trueman's frustration with conventional interface devices available to the violinist; footpedals seem crude and awkward as expressive instruments in comparison to the bowed string. By itself, the Rbow suggests a variety of kinds of physical interaction with electronic sound; moving the frog in various positions, which may require moving the entire body, and simply pressing the bow in various locations, all are effective ways of physically playing the Rbow. In this way, the Rbow transforms the violinist into a kind of dancer, and requires Trueman to modify his traditional violin technique. When played with the electric violin, this often creates an interesting technical conflict—certain techniques, while effective for the Rbow, may be useless for playing the electric violin, and vice-versa. Finding points of cross-section, where playing both instruments simultaneously is physically and musically fulfilling, is one of the fascinating challenges presented by the Rbow.

The Rbow also offers mapping flexibility; the sensor data can be interpreted in many ways, and these interpretations can change over the course of a performance, phrase, or even within a gesture. One particularly compelling mapping attaches virtual shakers to the bow, encouraging the performer to shake and gyrate in various ways to control the energy and resonances of the shakers. In combination with analysis of the violin's audio signal, the Rbow makes for an effective controller of both physical models and realtime granular delay techniques; Trueman makes extensive use of the PeRColate toolkit in MSP which includes both of these kinds of synthesis and signal processing techniques.

The website for this paper includes an audio example of Trueman playing the electric violin with Rbow and granular sampling and a video example of the Rbow with virtual shakers.

### Sbass – the Sensor Bass

The Sbass interface is built on a 5 string electric upright string bass. To the basic instrument an array of pickups has been added, including a contact microphone mounted under the hair of the tip of the bow. A small mouse touch-pad is mounted under the fingerboard offering two axes of continuous control in performance, and several extra buttons.

Since much of Bahn's playing is pizzicato, rather then focusing on the bow as the primary dynamic interface, sensors are mounted to the body of the instrument itself. These include several slide sensors, force sensitive resistors, turn "pots," and, a biaxial accelerometer. The tactile sensors are placed in such a way that they can be easily manipulated while playing, the accelerometer allows slight movement and tilt of the bass to control dynamic aspects of the sound. Using a micro-controller mounted on the side of the bass, the signals from these sensors are scaled into MIDI continuous control data affecting performance parameters of the MAX/MSP performance patch running on the main performance computer.

Bahn's MAX/MSP "patch" also makes extensive use of the PeRColate toolkit emlpoying numerous delay and granulation routines, physical modeling instruments, several different kinds of filtering, numerous palettes of sampled sounds, a mixing/routing console, and an algorithmic mixing routine drawing upon a large collection of composed sounds grouped in textural categories. The interactive computer environment is designed to maximize flexibility in performance to generate, layer and route musical material with the same improvisational freedom as he has developed with his string bass.

An aim in creating this interface was to enable Bahn to take his electro-acoustic music out of the studio and into a wide range of performance contexts. The configuration of the sensors and the computer performance interface is constantly changing and developing in a way analogous to the musical development of an improviser from performance to performance.

The website for this paper includes a series of excerpts from Bahn's new CD "r!g," which is a set of live solo "Sbass" improvisations.

## SPHERICAL SPEAKER ARRAYS and SENSAS

Essential to the development of subtle gestural performance interfaces are equally responsive sonic displays; they are of central importance in the feedback loop between physical gesture and sonic response. The sonic display must reinforce the nuance of physical gesture and offer localized sonic feedback for the performers on stage.

### Making Electronic Music More Intimate

As we integrated electronics and computation into our improvisations, our conventional sound system grew, eventually completely obscuring our acoustic beginnings. In our most recent season, we replaced our P.A.-style system with a set of six spherical speaker arrays of various sizes, including three 14-inch spheres, a 12-inch sphere, an enormous 22-inch sphere (Bubba; described further below), and an 8-inch tweeter-ball. These speakers, strewn about the stage in various configurations, function much like instrumental sources and create a sound field somewhat similar to a conventional chamber ensemble. The spheres localize our sounds, providing distinct points on stage for listeners and performers to grasp, yet also fill spaces and encourage listeners to walk among us; the typical plane of separation created by stage and P.A. system is non-existent. Consequently, these speakers render the concept of "monitoring" irrelevant; there is no need to create a "monitor mix" since the speakers and room do it automatically. Finally, these speakers engage the reverberant qualities of the performance spaces they are in, allowing the electronic sound to blend with acoustic sources (which is essential when we play with guest musicians) and making it unnecessary to add artificial reverberation (though it is sometimes interesting to do so in any case).

We find that this sound system drastically affects the way we play our electronic instruments, encouraging us to play softly and explore spare textures. It also feels familiar, reminding us (distantly) of our earlier, more "acoustic" improvisations and has forced us to see these display devices as an inseparable component of our extended instrumental and ensemble set-ups; they are as much a part of our instruments as the strings and bridges are.

**Sensor/Speaker Arrays: BoSSA and Bubba**

Given the instrumental qualities of spherical speaker arrays, it makes sense to actually imagine them as instruments. The Bowed-Sensor-Speaker-Array (BoSSA), constructed by Trueman, combines a 12-channel spherical speaker array with a variety of sensors inspired by the physical interface of the violin. This instrument, which was our first Sensor-Speaker-Array (SenSA), has been used in performance many times and is undergoing constant refinement. BoSSA suggested the possibility of a new kind of electronic chamber music and, with this in mind, Bahn constructed an enormous 22-inch 12-channel spherical speaker (Bubba) and a matching (much smaller) sensor-ball (the Bubba-Ball) to "play" Bubba.

In combination, BoSSA and Bubba form a compelling, if somewhat bizarre, ensemble. Both SenSAs reorient our relationship with electronic sound and convey a strong sense of physicality in performance. No longer detached from the sound source, we often feel as though we hold the sound in our hands. The experience is similar to performing with an acoustic instrument and the effort required can be equally exhausting. We are particularly impressed by the sense of presence and intimacy they provide. Given the lack of any direct acoustic sound source, we have freedom to redefine the mappings from sensor to synthesis/signal processing parameter on the fly, which in turn transforms our physical relationship with the instrument; in a sense, instrument design itself becomes an aspect of performance (we don't have to go back to the shop to alter the "feel" of our instruments; we do it in realtime).

The website for this paper includes images of BoSSA and Bubba and video of a performance of the "Lobster Quadrille," for BoSSA.

**COLLABORATIVE PERFORMANCES**

A full "interface" performance involves not only solo and duo performances by Trueman and Bahn, but collaborative compositions and performances with others employing similar interactive technologies. This includes interactive video performances with Nick Fortunato and Erin Seymour, interactive dance performances with Tomie Hahn, performances with Wacom-Tablet/spoken-word artist Monica Mugan, and collaborative improvisations with Perry Cook.

*Streams*

"Streams" is an interactive sonic context for live performance developed by Curtis Bahn and dancer/ethnomusicologist Tomie Hahn. Wearing a sensing device developed by Bahn, Hahn freely navigates a virtual sonic geography consisting of synthetic sounds and non-linear poetry. Through her movement, she is able to negotiate and control all aspects of the sonic structure of this virtual soundscape. With each gesture "Streams" recalls bodies of water and land, technology, a flow of information, transmission, and liquid states. Through technology, the performance toys with the ephemeral quality of sound and the physical memory of time, sonic space, and sensory experience.

Sensors on Hahn's hands each sense pressure and 2 axis of tilt, making 6 "channels" of continuous control information available in performance. The data from these sensors is sent to a micro-controller where it is translated into MIDI messages and broadcast to the computer offstage. A MAX/MSP patch on the computer maps her movements into various sampled sounds and DSP algorithms, sending them to an 8 channel audio system.

Bahn/Hahn, with a fifteen year history of collaborative performance, propose that the nature of the technologies employed in "Streams," fundamentally changed aspects of their collaboration regarding movement and sound composition. Rather then structuring time, as in traditional/historical dance/music collaborations, the conception of "Streams" was based on "composing the body." In this process, physical attributes of the dancer's movement vocabulary were analyzed to extract particularly salient and meaningful gestures. A parameter-mapping system was devised allowing the dancer to freely navigate and layer sonic elements to construct a complex texture.

In "Streams," Hahn draws on over thirty years of experience in Japanese traditional and experimental dance. The tradition of Japanese dance is brought to a contemporary expressive moment with the sensor interface tapping the "site" of her personal embodied knowledge of this tradition. Just as the senses are an individuals' "interface" to the world, technological interfaces can integrate vocabularies of ancient traditions of performance with the contemporary body.

The sonic palette employed in "Streams" draws from a combination of real-time digital signal processing, physical modeling synthesis algorithms (again from PeRColate) and stored sound samples of text. At the heart of the computer performance system is a digital model of the filtration characteristics of the vocal tract, all other sounds are passed through this sonic model evoking the image that, through her movements, the dancer "speaks" the music. Other sound sources are drawn from the technique of physical modeling synthesis, which, when paired with physical movement sensors, provides a particularly rich and evocative sonic landscape. The dancer also provides data controlling the construction of an algorithmic, non-linear text drawing from words relating to dreams, "flow," and communication.

The time-structure of "Streams" is not specified, the dancer is free to explore the sounds according to her feelings in the moment. However, drawing upon a basis of highly specified algorithmic compositional processes, neither is it improvised. The composition embodies physical mappings as they relate to the specific dancer's movement vocabulary, and the sound-world of the composition creating a highly personal and moving site-specific statement; "a personal sonic geography."

The website for this paper includes images and streaming video of a performance of "Streams."

**SPeaPer Interface, the "Sensor Speaker Performer"**

As an outgrowth of investigations in the use of Sensor/Speaker Arrays for live performance with "interface," we became interested in the idea of mounting not only sensors, but also small speakers on a live performer. The SPeaPer (Sensor/Speaker Performer) wireless interface builds on the sensor design for "streams," adding a small wireless amplifier and two arm-mounted coaxial speakers. Two channels of the computer generated sound are sent to Tomie's arms (discreet information to the left and right). The remaining 6 channels of audio are sent to 3 stereo spherical speaker arrays on stage.

The dance/movement composition "*pikapika*" (meaning "twinkling" in Japanese) sonifies her movement with sampled sounds of machinery, gears and rachets. *Pikapika*--a character influenced by *anime* and *manga*, (Japanese pop animation and comics) embodies movements from *bunraku* (puppet theater). The concept of the sonic punctuation of movement is drawn directly from the *bunraku* musical tradition, however, the actual sounds are not drawn from bunraku musical vocabulary. The sounds emanate from her body-mounted speakers create a new sort of audio "alias" for her character; a sonic mask.

The website for this paper includes images and streaming video of a performance of "Streams."

**SUMMARY**

We see an accelerating paradigm switch in computer music from being a field primarily concerned with composition to one equally involved with performance. In the social activity of ensemble performance, musical interface design must be concerned with the communication of gesture and sonic nuance. We see the importance of this not only in terms of the physical communication between members of a musical ensemble, but also for the communication of electronic music to an audience. The inclusion of sound reinforcement in musical interface designs also provides important aural and tactile feedback cues to the performers allowing a more intimate interaction with electronic music production. In this sense, our "interfaces," our "instruments," include the entire systems from physical sensing to computer algorithms to sonic display devices. This decreases the generality of the music we produce; it is not a generic "tape" that can be reproduced on any sound system. It is rather an idiosyncratic sonic performance installation for a unique family of instruments.

In this instrumental model, our goal is to privilege and enhance the human interaction and social context of music making. We do not necessarily see the systems as "interactive," in the sense that "interaction" implies a sort of musical autonomy for the technology, but rather, a goal may be to have an electronic music interface as "interactive" as an acoustic piano or a violin. The difference between the acoustic instrument and an advanced interface such as this is in the ability to reconfigure and alter the many parameters of control and their significance in performance. This dynamic mapping extends the potentials of the instruments into the realm of a non-linear composition. These dynamic "composed instruments" are defined by the capabilities of the acoustic and "virtual" aspects of the interfaces, the collection of resources for computer sound production and processing, and the sonic display of the performance. In our work with "interface" we choose to keep these potentials as freely organized as possible, allowing the exploration of their potentials in improvisatory ensemble performance.